\documentclass[superscriptaddress,twocolumn,amsmath,amssymb,aps,prb,floatfix,showkeys,
]{revtex4-2}
\usepackage{graphicx}
\usepackage[colorlinks = true,
            citecolor = blue,
            linkcolor = blue,
            urlcolor = blue]{hyperref}
\usepackage[all]{hypcap}
\usepackage{xcolor}
\usepackage{gensymb}
\usepackage{float}
\usepackage{setspace}
\usepackage{lineno}

\newcommand{\HAM}{Institute of Nanostructure and Solid State Physics (INF), University of Hamburg, Jungiusstraße 11, 20355 Hamburg, Germany}

\begin{document} 
\title{Reciprocity of Magnetism and Nanostructure Growth}
\author{Felix Zahner}\email[]{fzahner@physnet.uni-hamburg.de}\affiliation{\HAM}
\author{Kirsten von Bergmann}\affiliation{\HAM}
\author{Roland Wiesendanger}\affiliation{\HAM}
\author{André Kubetzka}\email[]{kubetzka@physnet.uni-hamburg.de}\affiliation{\HAM}
\date{\today}

\begin{abstract}
    The growth of thin films and nanostructures is a fundamental process for constructing both model-type systems and nanoscale devices, where performance and functionalities can be controlled by the choice of parameters. For magnetic systems the crystal structure, material composition, size, and shape determine already most of the magnetism-related properties. Here, we investigate the reciprocal route, which is controlling nanostructure growth via magnetism. To this end we deposit Co on a uniaxial antiferromagnetic surface above and below its Néel temperature $T_{\text{N}}$. Growth above $T_{\text{N}}$ results in the formation of roughly hexagonal islands, reflecting the surface symmetry.
    For growth below  $T_{\text{N}}$ we find quasi-one-dimensional Co nanostructures along distinct crystallographic directions, which signal the local antiferromagnetic domain orientation.
    Our findings demonstrate the feasibility of controlling growth via magnetism and the necessity to take magnetism-related effects into account for growth on magnetic surfaces.
\end{abstract}
\maketitle
\clearpage

\section*{Main}
The epitaxial growth of nanostructures on crystalline substrates is a well-studied process which allows for the preparation of self-organized structures for fundamental research as well as spintronic~\cite{Baltz2018} and computing applications~\cite{Grollier2020,Everschor2024}.
A large variety of different shapes and functionalities of nanostructures can be achieved by the choice of materials, the surface symmetry, and  external growth parameters such as surface temperature and evaporation rate~\cite{Michely1993,Roder1993}. By controlling the size and shape of these nanostructures, chemical reactivity as well as the electronic and magnetic properties can be tuned. In the context of growth, kinetic processes can compete with thermodynamics and the interplay of many different diffusion processes with their respective activation energies can govern the resulting morphology. At low temperatures and hindered edge-diffusion a "hit-and-stick" mechanism can lead to fractal growth~\cite{Michely1993}, while with increasing temperatures additional diffusion pathways open until finally thermodynamically stable, more compact two-dimensional shapes are formed. One-dimensional (1D) structures including atomic chains can be prepared as thermodynamically stable states at substrate step edges~\cite{Gambardella2002} or as metastable configurations as a result of anisotropic atom diffusion on low-symmetry surfaces~\cite{Roder1993,Hammer2003,Menzel2012,Ferstl2016}.

For magnetic systems, structural control allows, e.g., to select specific spin textures~\cite{Hagemeister2016,Menzel2012,cortes-ortuno_nanoscale_2019}, to increase the crystalline anisotropy~\cite{Gambardella2005}, and tune the superparamagnetic blocking temperature~\cite{Bode2004}. The inverse case, where a magnetic effect is exploited to control the structure, is rare and so far limited to the use of external magnetic fields. For instance, in solution nano-particles can be assembled by magnetic fields into chain-like structures~\cite{Niu2004,Zhang2009}. Similarly, the interaction of the stray field from magnetic domain walls with hematite particles in solution was used for the first magnetic imaging technique developed by F.\,Bitter~\cite{Bitter1931,Bitter1932}. 
In the case of growth from the gas phase on a solid substrate, it has been shown that the atomic structure of magnetic films can be altered by external magnetic fields, giving rise to crystalline anisotropy~\cite{Chen1992} or exchange bias~\cite{Yuan2022}.
Except for such examples, the growth of magnetic films and nanostructures is considered to be unaffected by magnetic interactions, either because the nanostructures are grown above the magnetic ordering temperature or because magnetic interactions are typically 2--3 orders of magnitude weaker then adhesion and bonding energies. Nevertheless, J.\,Hafner and coworkers predicted a strong influence of the magnetic state of Mn and Fe monolayers onto diffusion barriers and even adsorption sites for Mn and Fe adatoms~\cite{Spisak2004,Dennler2005}, calling the standard view into question. Along these lines we have shown in a previous experimental work that a row-wise antiferromagnet (RW-AFM) can restrict adatom motion to one dimension, when this motion is initiated at low temperature by local voltage pulses from a scanning tunneling microscope (STM) tip~\cite{Zahner2025}. Consequently, this finding suggests that the control of nanostructure growth may be realized by magnetic spin textures.

Here, we use the antiferromagnetic fcc Mn layer on Re(0001) to investigate the submonolayer growth of Co above and below $T_{\text{N}}$ of the RW-AFM state. For growth at $300$\,K we find roughly hexagonal reconstructed Co monolayer islands. Upon lowering the temperature below $T_{\text{N}}$, these compact two-dimensional Co islands generate the formation of antiferromagnetic domain walls in their vicinity. For growth at $80$\,K the Co is also reconstructed, but predominantly forms 1D-nanostructures with aspect ratios on the order of 5--10. We demonstrate a strong link between the direction of 1D-nanostructure growth and the orientation of the antiferromagnetic domains, enabling a Bitter-technique-like imaging of antiferromagnetic domains.

\section*{Interplay of magnetism and growth}
\begin{figure}[h]
    \centering
    \includegraphics[width=0.95\linewidth]{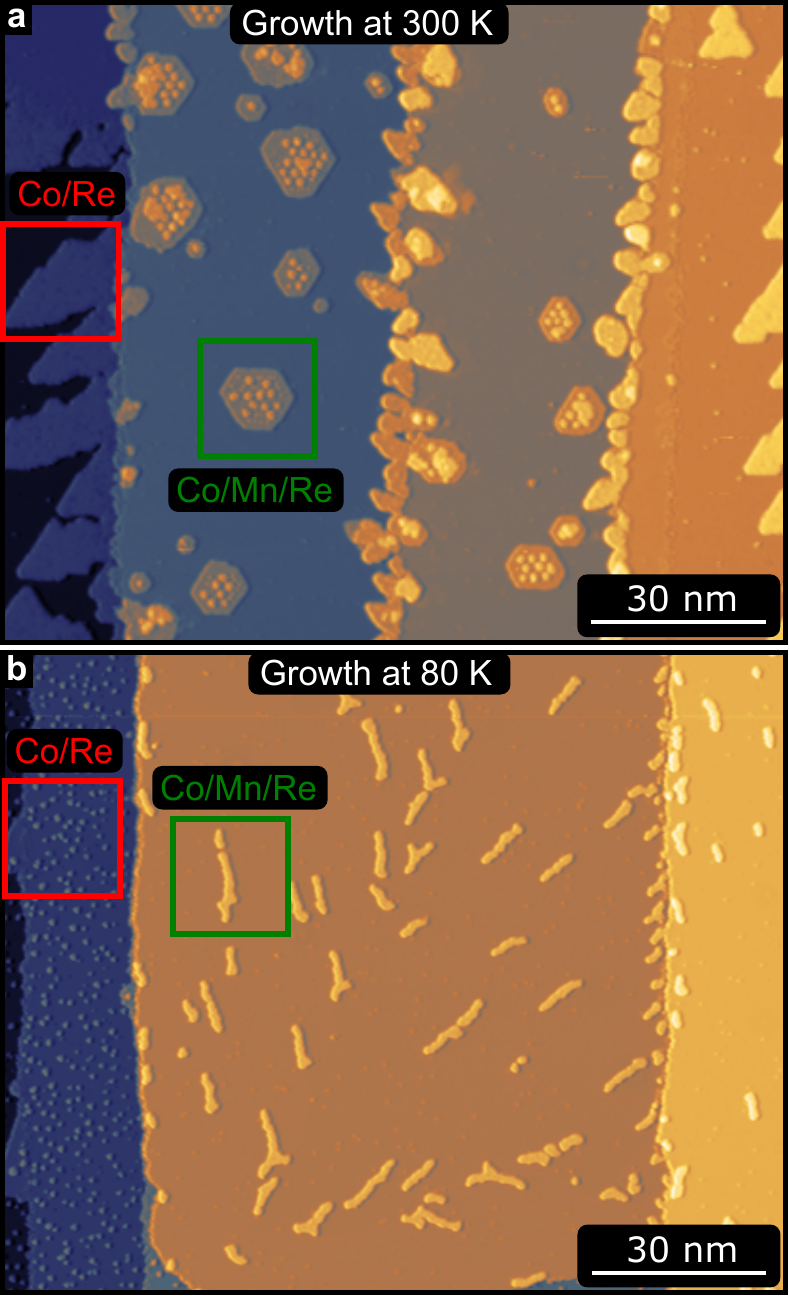}
    \caption{\textbf{| Magnetism affects growth. a,}~Partially differentiated STM image of Co deposited at room temperature, showing hexagonal and triangular Cobalt islands on the Mn monolayer and the bare Re(0001) substrate, respectively. $U=200$\,mV, $I=500$\,pA. \textbf{b,}~Partially differentiated STM image of Co deposited at 80\,K. Islands grow with a large aspect ratio on the Mn ML. Orientation depends on the underlying orientational domain of the RW-AFM state. Co forms small clusters on the bare Re(0001) surface. $U=500$\,mV, $I=500$\,pA. Both images were measured at 4.2\,K.}
    \label{fig1}
\end{figure}

To investigate the impact of magnetism on growth, we deposit Co onto an antiferromagnetic substrate above and below its Néel temperature $T_\text{N}$. We choose a single layer of fcc-stacked Mn on Re(0001)~\cite{Spethmann2020} as a substrate, a well-characterized system where we found tip-induced 1D-motion of adatoms resulting from its RW-AFM state~\cite{Zahner2025}.
Due to the hexagonal atom arrangement, in combination with this uniaxial magnetic state, three orientational domains of the RW-AFM state exist, which are separated by superposition domain walls~\cite{Spethmann2021}.
While the Néel temperature ($T_\text{N}$) estimated using Monte Carlo methods, based on density functional theory, is $T_\text{N}=160\pm 5$\,K~\cite{Gruenebohm2022,Spethmann2021}, a value of $T_\text{N}=75\pm 5$\,K was derived from a recent momentum microscopy study~\cite{Elmers2023} (see methods for more information).

When Co is deposited onto the Mn monolayer at room temperature (i.e., when the system is far above the magnetic ordering temperature), it grows in the form of roughly hexagonal monolayer islands, as expected due to the symmetry of the surface, with the island edges running along the close-packed row directions, see Fig.\,\ref{fig1}a. Due to the large lattice mismatch of Co and the substrate, strain is relieved by an irregular hexagonal reconstruction (see Extended Data Fig.\,1). This superstructure is decorated with additional Co clusters on top. On areas where the Re surface is exposed, the Co forms islands of approximately triangular shape, see red box in Fig.\,\ref{fig1}a.

When Co is deposited at  $\approx80$\,K, the diffusion length of the Co atoms is strongly reduced on the Re substrate, resulting in a high density of small clusters, see red box in Fig.\,\ref{fig1}b. On the Mn monolayer the Co predominantly forms 1D-nanostructures with the long axes in specific crystallographic directions. Because of this drastic change of the morphology we conclude that the growth temperature is now below the Néel temperature, i.e., during the Co growth the Mn monolayer is antiferromagnetically ordered.
While 1D-nanostructures in close vicinity are typically oriented in the same direction, in other sample regions the long axis is rotated by about $\pm 120^{\circ}$.

\begin{figure}[h]
    \centering
    \includegraphics[width=0.95\linewidth]{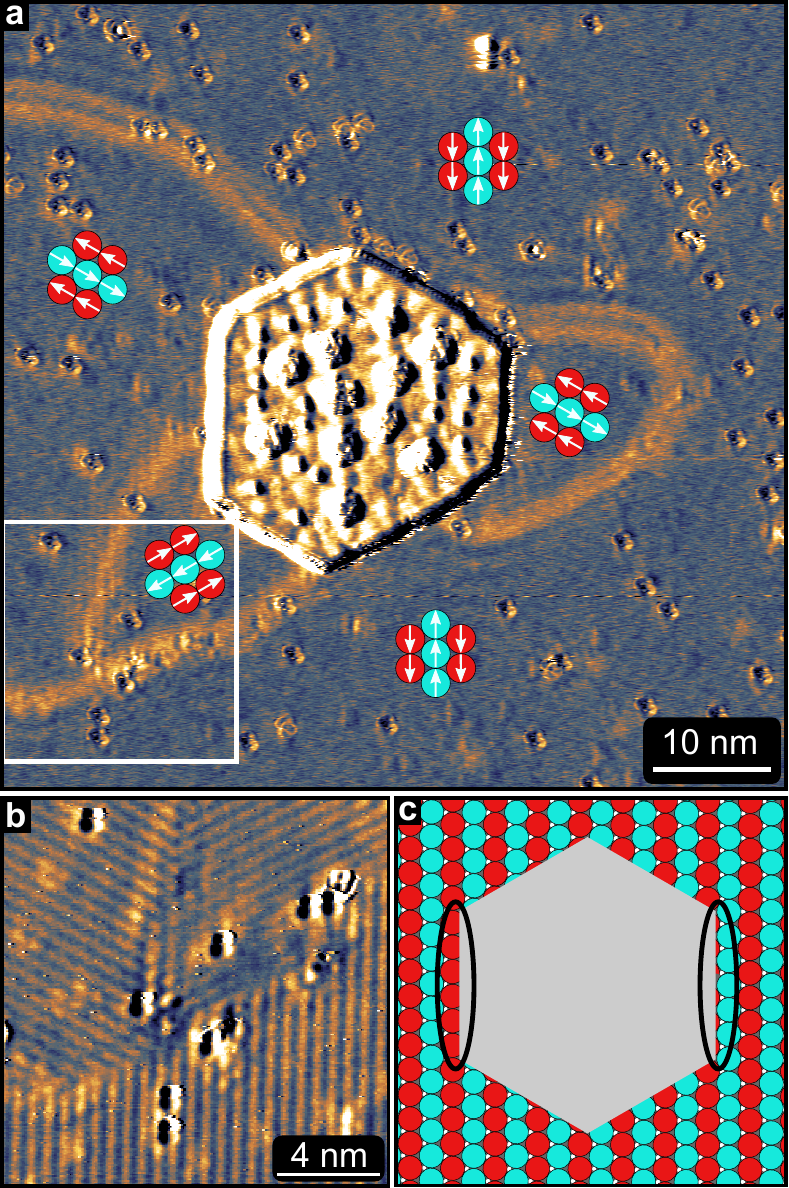}
    \caption{\textbf{| Co islands disturb the magnetic state. a,}~d$I$/d$U$-map of a hexagonal Co island on the Mn monolayer. The bright orange lines indicate the domain wall positions. RW-AFM domain orientations are marked by the red-cyan sketch (not to scale). $U=10$\,mV, $I=3$\,nA. \textbf{b,}~SP-STM image of the region marked by the white rectangle in a. $U=10$\,mV, $I=5$\,nA. \textbf{c,}~Sketch of the possible orientations of rows of parallel spins at the edges. The edges with a parallel alignment of spins (marked by black ellipses) are unfavorable. All images were measured at 4.2\,K.}
    \label{fig2}
\end{figure}

To understand the mutual interactions of magnetism and growth, we have used spin-polarized (SP)-STM at 4.2\,K~\cite{Wiesendanger2009,vonBergmann2014}. The positions of domain walls in the Mn monolayer can be identified in measurements with a non-magnetic tip, due to their distinct structural and electronic properties~\cite{Spethmann2021,Zahner2025}. We find that in the vicinity of the hexagonal Co islands, which were grown above the Néel temperature, small orientational domains form at some of the edges when cooling below the magnetic ordering temperature, see Fig.\,\ref{fig2}a and Extended Data Fig.\,1. For the pristine uncovered Mn monolayer, domain walls are observed much less frequently~\cite{Spethmann2020,Spethmann2021}, which suggests that the Co islands are responsible for the nucleation of these small domains.

The orientation of the antiferromagnetic domains can be identified in SP-STM measurements, see Fig.\,\ref{fig2}b, where the RW-AFM state is imaged as stripes with a spacing of two atomic row distances~\cite{Spethmann2020}. Based on this measurement, the antiferromagnetic domains in Fig.\,\ref{fig2}a have been labeled by small red and cyan icons. We find that the parallel rows of spins (e.g.\ the cyan atomic rows) of the antiferromagnetic domains always have an angle of $\pm 120^{\circ}$ with respect to the edges of the adjacent Co island. Fig.\,\ref{fig2}c shows a sketch of a hexagonal island on a single domain of the uniaxial RW-AFM state. The Co island edges are oriented along close-packed row directions. Edges where the RW-AFM state is oriented in a way that the spins are aligned parallel to the island edges ($\uparrow\uparrow$-edges) are marked by black ellipses. For the other edges, the spins are alternating along the edge ($\uparrow\downarrow$-edges). Our experiments demonstrate that the Co islands on the Mn monolayer exclusively have the $\uparrow\downarrow$-edges. The $\uparrow\uparrow$-edges are avoided by introducing domain walls as seen in Fig.\,\ref{fig2}a and Extended Data Fig.\,2, thereby forming favorable orientational domains in the vicinity of each hexagonal Co island. We will refer to this as edge-magnetism-frustration. This is similar to the selection of specific orientational domains of the RW-AFM state previously observed for step edges~\cite{Spethmann2021} and local strain~\cite{Saxena2024}.

\begin{figure}[h]
    \centering
    \includegraphics[width=0.95\linewidth]{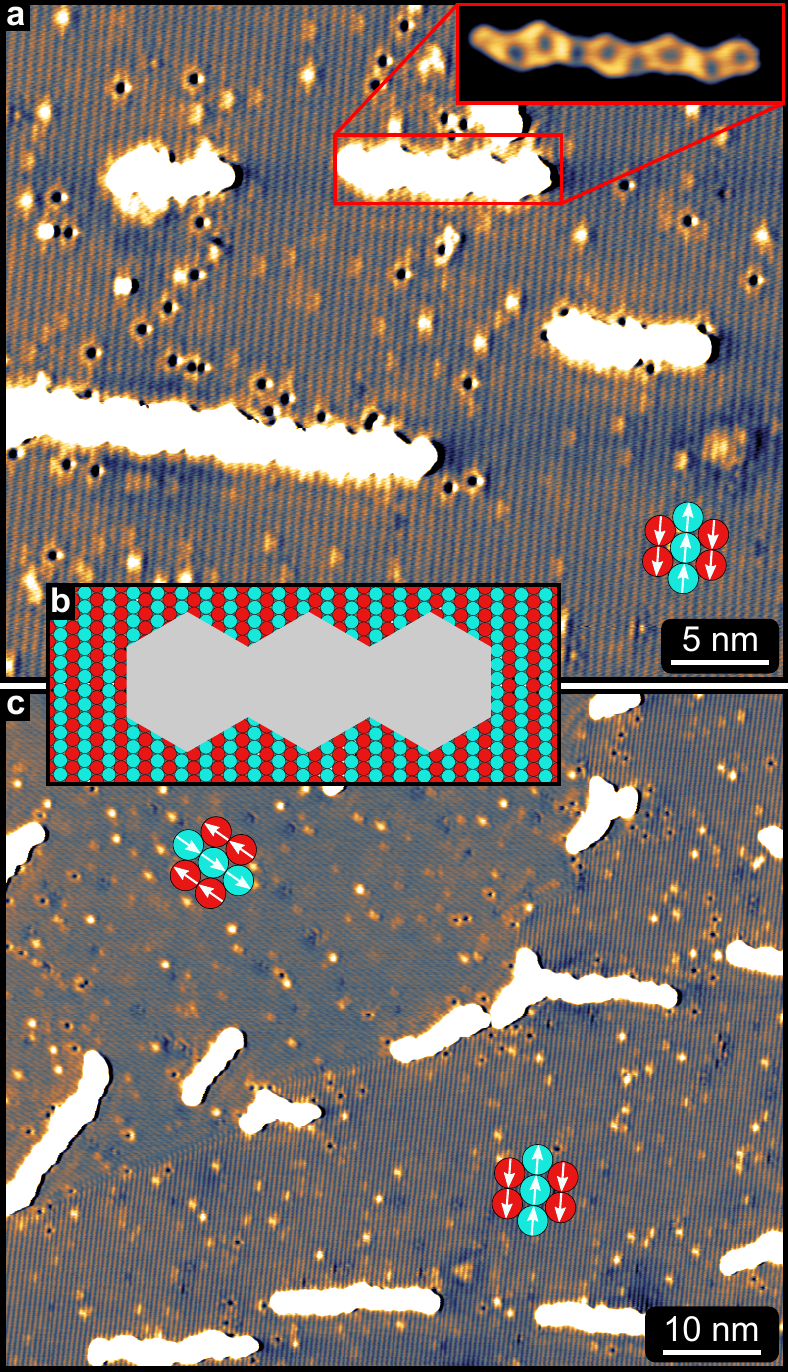}
    \caption{\textbf{| 1D-nanostructures grown at 80\,K. a,}~Partially differentiated SP-STM image of elongated Co islands grown at 80\,K. The rows of the RW-AFM state intersect the long axis of the islands at a  $\approx 90\degree{}$ angle. Inset shows the topography of such a 1D-nanostructure. $U=10$\,mV, $I=7$\,nA. \textbf{b,}~Sketch of the orientation of rows of parallel spins at the edges of an approximated 1D-nanostructure. \textbf{c,}~Partially differentiated SP-STM image of Co 1D-nanostructures on two orientational domains of the magnetic state, as the quantization axis of the antiferromagnetic state rotates by 120\textdegree~the long axis of the nanostructures rotates in the same way, highlighting the coupling to the magnetic state. $U=10$\,mV, $I=4$\,nA. All images were measured at 4.2\,K.}
    \label{fig3}
\end{figure}

In the case of the Co 1D-nanostructures grown onto the Mn monolayer at $80$\,K, we observe that the stripes of the RW-AFM state are always perpendicular to the long axis of the Co structures, see Fig.\,\ref{fig3}a,c. This closer view also reveals that the 1D-nanostructures consist of smaller roughly hexagonal building blocks, see inset in Fig.\,\ref{fig3}a, reminiscent of the irregular hexagonal superstructure in the larger monolayer islands (see Extended Data Fig.\,1) and equally originating from the lattice mismatch driven local strain relief. The shape of the 1D-nanostructures can be approximated by a chain of hexagons, see sketch in Fig.\,\ref{fig3}b, which also serves to illustrate the fact that the zig-zag shaped long edges have the same edge-magnetism relation as the hexagonal Co islands, i.e., the Mn spins alternate along these edges ($\uparrow\downarrow$-edges); the unfavourable $\uparrow\uparrow$-edges, where Mn spins are parallel along the Co edge are present only at the two ends of the Co 1D-nanostructures.
At the positions of domain walls the shape of the Co nanostructures can deviate from the strictly 1D-nature, see Fig.\,\ref{fig3}c. It is worth to note that the Co structures exhibit a preference to grow at domain walls, see Fig.\,\ref{fig1}b.

\begin{figure}[h]
    \centering
    \includegraphics[width=0.95\linewidth]{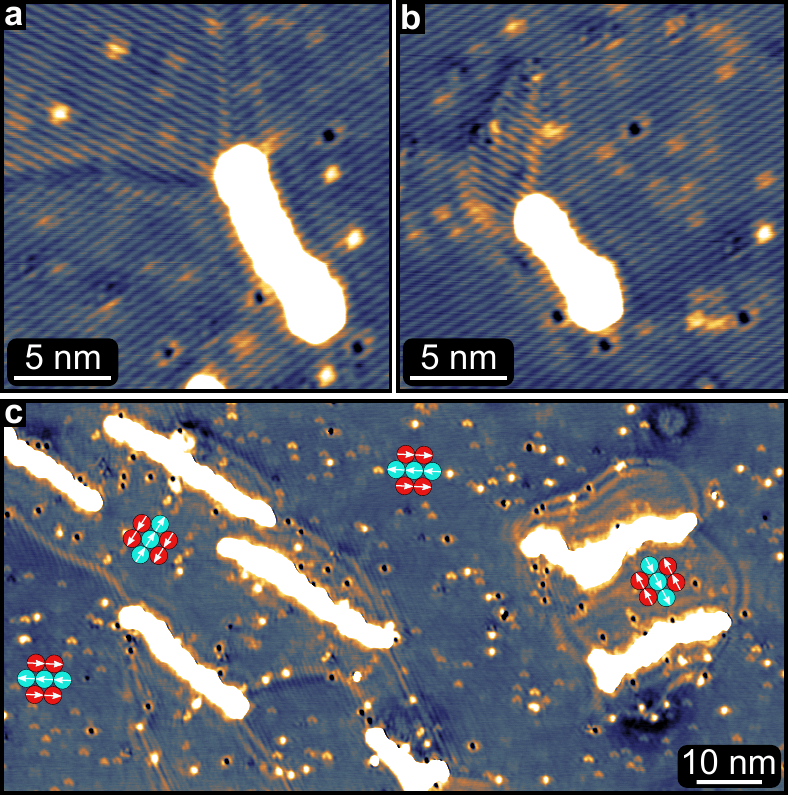}
    \caption{\textbf{| Interaction of magnetic state and nano-structure. a,}~SP-STM image of the magnetic state around an elongated Co island. The domain wall is positioned in such a way, that the short edge of 1D-nanostructure is exposed to the other orientational domain. $U=5$\,mV, $I=5$\,nA. \textbf{b,}~SP-STM image of the magnetic state around a 1D-nanostructure. A domain at the short edge of the islands is visible. $U=5$\,mV, $I=5$\,nA. \textbf{c,}~STM image of an area where the main domain of the terrace has changed after the growth of the islands, with the elongated islands locally pinning the orientational domain present during growth. Domain orientation is marked by red-cyan sketch. $U=5$\,mV, $I=5$\,nA. All images were measured at 4.2\,K.}
    \label{fig4}
\end{figure}

In some cases, we observe domain wall pinning or even a small enclosed domain at the ends of the 1D-nanostructures, see Fig.\,\ref{fig4}a,b. This indicates that there is an energy gain associated with the avoidance of the $\uparrow\uparrow$-edge, supporting the previous interpretation sketched for the hexagonal Co island in Fig.\,\ref{fig2}c. 
The strong link between the 1D-nanostructures and the orientational domains of the RW-AFM state can be seen in Fig.\,\ref{fig4}c. Here a domain wall must have moved across the terrace after the growth of the 1D Co nanostructures, leading to a change of the domain pattern, see Fig.\,\ref{fig4}c and Extended Data Fig.\,2. In this sample area we observed small closure domain walls around groups of 1D-nanostructures with the same long axis orientation, demonstrating the clear connection between 1D-nanostructure and antiferromagnetic domain orientation.

\section*{Kinetics vs. thermodynamics}
In a previous investigation we found that at $4.2$\,K Co atoms move exclusively along the $\uparrow\uparrow$-rows of the RW-AFM state, when their motion is initiated by a local voltage pulse from an STM tip~\cite{Zahner2025}. Whereas these directly kicked atoms performed long jumps of up to 10\,nm, single site jumps of Co atoms further away from the tip showed the same strict movement direction. This atom hopping is reminiscent of single site jumps within a thermally driven random walk. It is therefore tempting to predict a temperature regime below $T_\text{N}$, where fully kinetically-driven growth leads to atomic Co chains oriented along the diffusion direction, as for instance observed on low-symmetry or reconstructed surfaces such as fcc(110) or Ir(001)-(5$\times$1)~\cite{Roder1993,Hammer2003,Menzel2012,Ferstl2016}. However, this should result in Co chains oriented perpendicular to the Co 1D-nanostructures observed here. Actually, no strict 1D-growth can be expected, because in contrast to structurally anisotropic systems, on the RW-AFM state adjacent 1D diffusion channels have no spatial separation and thus atoms in adjacent rows can meet and form compact clusters despite the 1D diffusion.

With increasing temperature the diffusion should become more isotropic when hopping between the $\uparrow\uparrow$-rows becomes thermally more and more accessible. A 90$^\circ$ rotation of the easy diffusion direction at $80$\,K, however, can be ruled out, because this is not even a close-packed row direction. Thus, surface kinetics alone cannot explain the observed 1D-growth. On the other hand, we know that the hexagonal island shapes observed in Fig.\,\ref{fig1}a are close to a minimum energy shape, so thermodynamics alone also cannot explain the 1D-shapes grown at $80$\,K. However, other effects also need to be taken into account. 
First, our experiments demonstrate that the hexagonal islands prevent a single domain state due to unfavorable edge-magnetism orientations (see black ellipses in Fig.\,\ref{fig2}c), which induces additional domain walls. Second, both the hexagonal islands as well as the 1D-nanostructures show a roughly hexagonal reconstruction arising from the lattice mismatch of $9.4\%$. Indeed, small compact shapes appear to be the initial building block for the 1D-nanostructures, and some of these can be seen in Fig.\,\ref{fig1}b. We speculate that due to the strain they have an optimum size, and further growth is associated with an increased energy barrier. 

From the shape of the 1D-nanostructures alone, we can see that the subsequent growth is faster in the long direction (i.e at the short edges). This can, for instance, originate from faster Co atom diffusion along those edges with the preferred magnetization configuration as compared to those with edge-magnetism frustration, see black ellipses in Fig.\,\ref{fig2}c. This would lead to an accumulation of Co atoms at the short edges, increasing the probability of nucleating an additional hexagonal building block, resulting in the 1D-nanostructure direction observed experimentally. 
This scenario of an interplay of kinetic and thermodynamic aspects together with edge-magnetism frustration and strain effects could explain the observed 1D-nanostructure growth of Co on the antiferromagnetic surface of the Mn monolayer on Re(0001).

\section*{Antiferromagnetic Bitter technique}
\begin{figure*}
    \centering
    \includegraphics[width=0.8\linewidth]{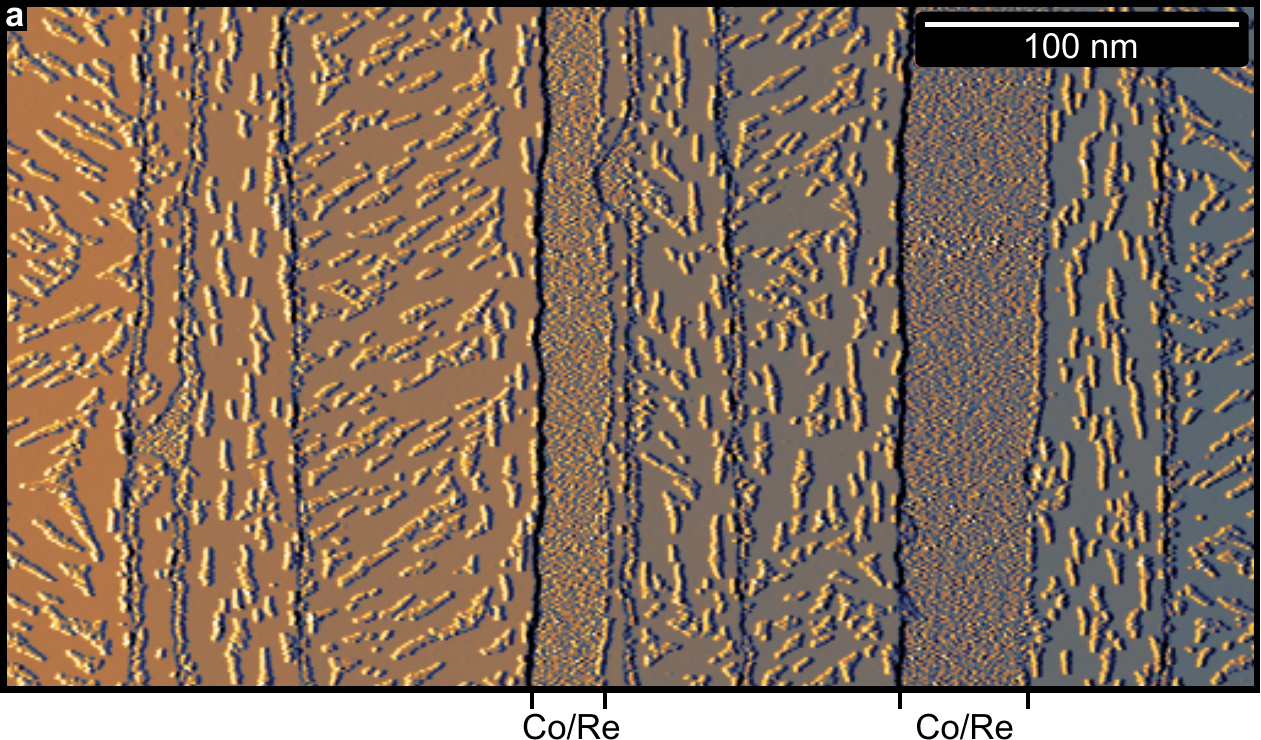}
    \caption{\textbf{| Mapping the large scale AFM domain structure. a,}~Partially differentiated STM image of 0.3\,ML Co coverage. The large scale configuration of elongated Co islands reveals information about the domain structure of the antiferromagnetic state. Strips of Co/Re areas are labeled. $U=500$\,mV, $I=300$\,pA. Image was measured at 78\,K.}
    \label{fig5}
\end{figure*}

In the original Bitter technique, magnetic nanoparticles in solution are used to decorate domain walls in ferromagnetic systems~\cite{Bitter1931,Bitter1932}. This process is based on the long-range magnetostatic interaction and various techniques have been used for the subsequent imaging~\cite{Hutchinson1965,Rice1991}. Our experiments show that in a similar way the anisotropic growth of Co on a RW-AFM surface can be utilized for the large-scale observation of the antiferromagnetic domain structure, without the need for spin-sensitivity. STM images of Co deposited below the Néel temperature show quasi-1D nanostructures with three principal orientations visible across the different terraces, see Fig.\,\ref{fig5}. Some nanostructures grow with 120\textdegree{} bends or in Y-shapes due to growth across or along domain walls. By inspecting the local morphology of the Co nanostructures we can thus directly observe the domain orientation, their size, and the location of domain walls between them. 

This new Bitter technique variant for antiferromagnets should be applicable to many systems with low-symmetry magnetic states giving rise to orientational domains: in addition to the RW-AFM, also a double-row-wise AFM~\cite{Kroenlein2018,Romming2018,Gutzeit2022}, short wavelength spin spirals~\cite{Ferriani2008,Romming2013,Khanh2020}, or skyrmion lattices~\cite{Heinze2011,Romming2013,Gutzeit2022} are ideal substrate candidates. In other systems, preferred nucleation at domain walls---similar to the original Bitter technique---might be exploited for domain wall imaging. 
Materials which grow pseudomorphically on an antiferromagnetic surface might develop other shapes, for instance diamond shapes which completely avoid $\uparrow\uparrow$-edges, or they might even prefer $\uparrow\uparrow$-edges or show only weak signs of broken symmetry. In any case, the drastic effect of the RW-AFM surface for Co growth demonstrates that magnetic spin textures are a means to control nanostructure growth. In general, the effect of the magnetism of the substrate onto growth is expected to be strongest for submonolayer coverages. However, deposited magnetic layers might inherit similar anisotropies themselves, or the initial 1D growth may serve as a template causing three-dimensional effects at higher coverages. 
A necessary requirement for this Bitter-like technique for antiferromagnets is that the material used for imaging needs to be sufficiently mobile below $T_\text{N}$ for the given antiferromagnetic surface. Compared to alternative techniques, the necessary experimental setup to investigate antiferromagnets is simple: any imaging technique that can identify a spatially varying anisotropy axis of the growth can be used. 

\section*{Conclusions}
We have investigated the growth of Co nanostructures on the row-wise antiferromagnet Mn/Re(0001) above and below its Néel temperature $T_\text{N}$. We find a striking reciprocity of magnetism and growth, with the hexagonal Co islands grown above $T_\text{N}$ causing magnetic frustration and the RW-AFM state channeling the Co atoms into 1D nanostructures below $T_\text{N}$, avoiding magnetic frustration. These 1D nanostructures signal the antiferromagnetic domain orientation, allowing magnetic imaging in a Bitter-like fashion. Our results demonstrate the feasibility to control growth via magnetism and the necessity to take magnetism into account for growth on low-symmetry magnetic surfaces.

\section*{Methods}
The experiments were done in an ultra-high vacuum (UHV) system using separate chambers for sample cleaning, metal deposition and STM measurements. Transfers between different chambers can be done without breaking the UHV.

\textbf{Sample preparation:} 
The Rhenium single crystal was cleaned by temperature cycles up to $1400$\,K in an oxygen atmosphere of $10^{-7}-10^{-8}$\,mbar. After a final flash annealing to $\approx 1800$\,K, Mn was deposited at a rate of $\approx 0.1$ atomic layers per minute on the Re crystal still at an elevated surface temperature of $\approx 100\degree$\,C. The Mn was evaporated from a pyrolytic boron-nitride (PBN) Knudsen cell of volume $2\,\mathrm{cm}^3$ at a temperature of 690\textdegree{}\,C. The sample was then cooled down to 80\,K or 300\,K, and subsequently Co was deposited at a rate of $\approx 1$\,ML/h from an e-beam evaporator using a 2\,mm Co rod. As the deposition was done with the sample in one of the STMs, the sample was measured in the same STM for the measurements at 80\,K while an in-situ transfer to another STM was needed for the measurements at 4.2\,K.

\textbf{SP-STM:}
The spin-polarized measurements were performed using a Cr-bulk tip, etched in 1M HCL solution. The tip was first cleaned via field emission on a W(110) crystal and subsequently sharpened using voltage pulses in the range of $4-6$\,V in addition to dipping the tip into the Mn/Re(0001) sample. Since we observed no magnetic signal on the Co nanostructures by SP-STM, their magnetic state is at this point unknown.

\textbf{Néel temperature:}
Concerning the Néel temperature of Mn/Re(0001), a discrepancy arises between our data and the results of H.J. Elmers and coworkers, who infer $T_\text{N}=75\pm 5$\,K from the temperature dependence of a band-splitting measured by momentum microscopy~\cite{Elmers2023}. Above $T_\text{N}$ the band-splitting, which is interpreted as an exchange-splitting, is finite and constant, indicating thermally fluctuating short range magnetic order up to at least $110$\,K. From the DFT magnetic parameters on the other hand, $T_\text{N}=160\pm 5$\,K can be estimated using Monte Carlo methods~\cite{Gruenebohm2022,Spethmann2021}. We have observed the growth of Co 1D-nanostructures up to a deposition temperature of 100\,K, indicating long-range magnetic order. Higher deposition temperatures were not investigated. Aspects which might contribute to this discrepancy are different levels of contamination, errors of the temperature measurements, or possibly an effective increase of $T_\text{N}$ due to the presence of Co at the early stages of growth.

\section*{Data availability}
The STM data are available from the corresponding authors upon reasonable request.

\section*{Code availability}
The STM data was analyzed with the open access software Gwyddion (http://gwyddion.net). 

\section*{Acknowledgment}
We acknowledge funding from the Deutsche Forschungsgemeinschaft (DFG, German Research Foundation) under project numbers 402843438, 408119516, and 418425860.

\section*{Author contributions}
F.Z. and A.K. devised the experiments, and prepared the samples. F.Z. and K.v.B. performed the STM measurements and together with A.K. analyzed the experimental data. The figures were prepared by F.Z. The manuscript was written by F.Z., A.K., and K.v.B., with all authors contributing.

\bibliography{References}

\newpage

 \renewcommand{\figurename}{Extended Data Fig.}
 \setcounter{figure}{0}

\begin{figure*}
    \centering
    \includegraphics[width=0.98\linewidth]{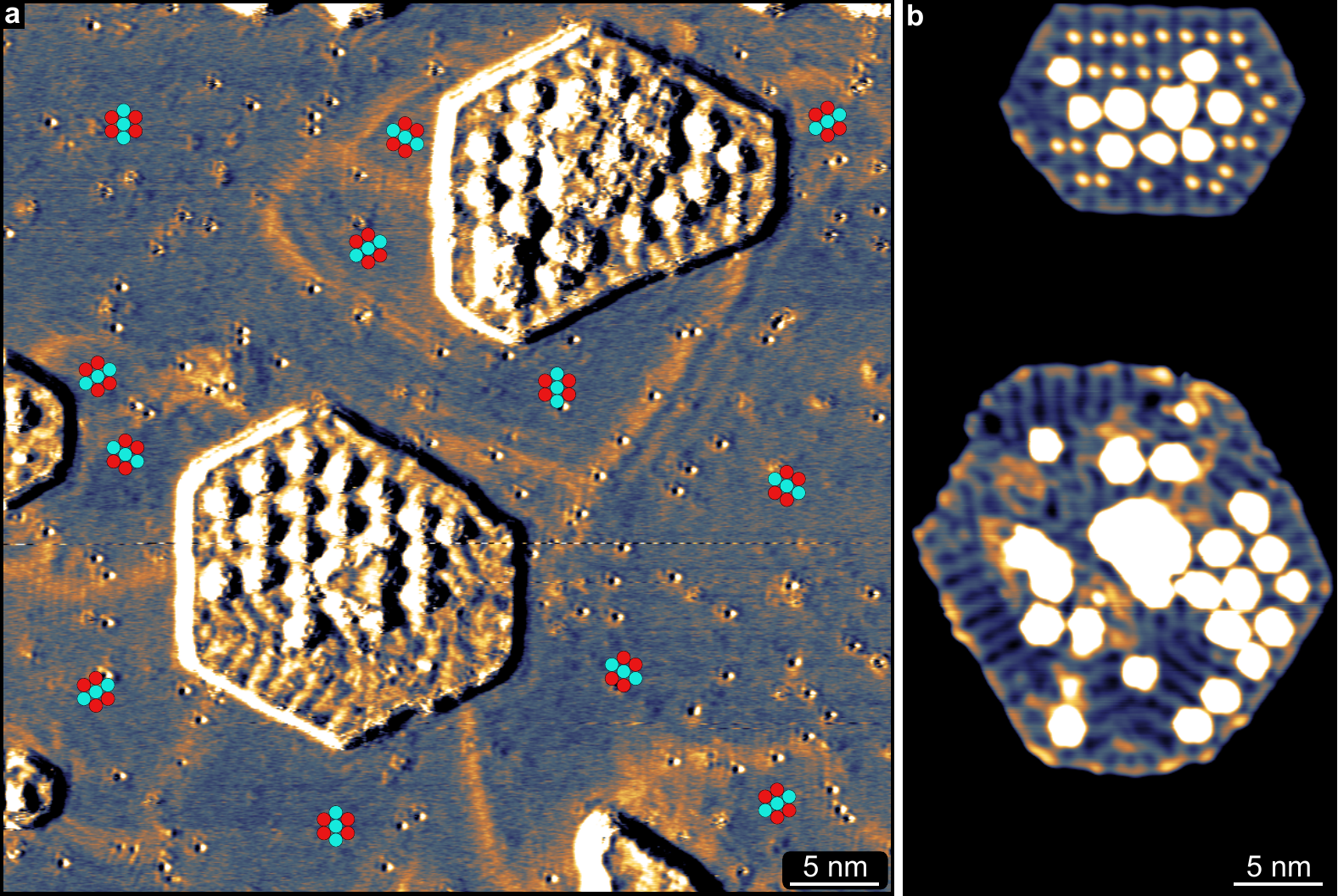}
    \caption{\textbf{| Domain structure affected by hexagonal Co islands. a,}~$\mathrm{d}I/\mathrm{d}U$-maps of Co islands deposited on Mn/Re(0001) at 300\,K. The bright orange lines are domain walls. Red-cyan sketches indicate the orientation of the domains. $U=10$\,mV, $I=3$\,nA. \textbf{b,}~STM image of two different Co islands. The color-map contrast is adjusted to only show the Co surface. $U=50$\,mV, $I=2$\,nA. Both images were measured at 4.2\,K.}
    \label{fig-Ex2}
\end{figure*}

\begin{figure*}
    \centering
    \includegraphics[width=0.98\linewidth]{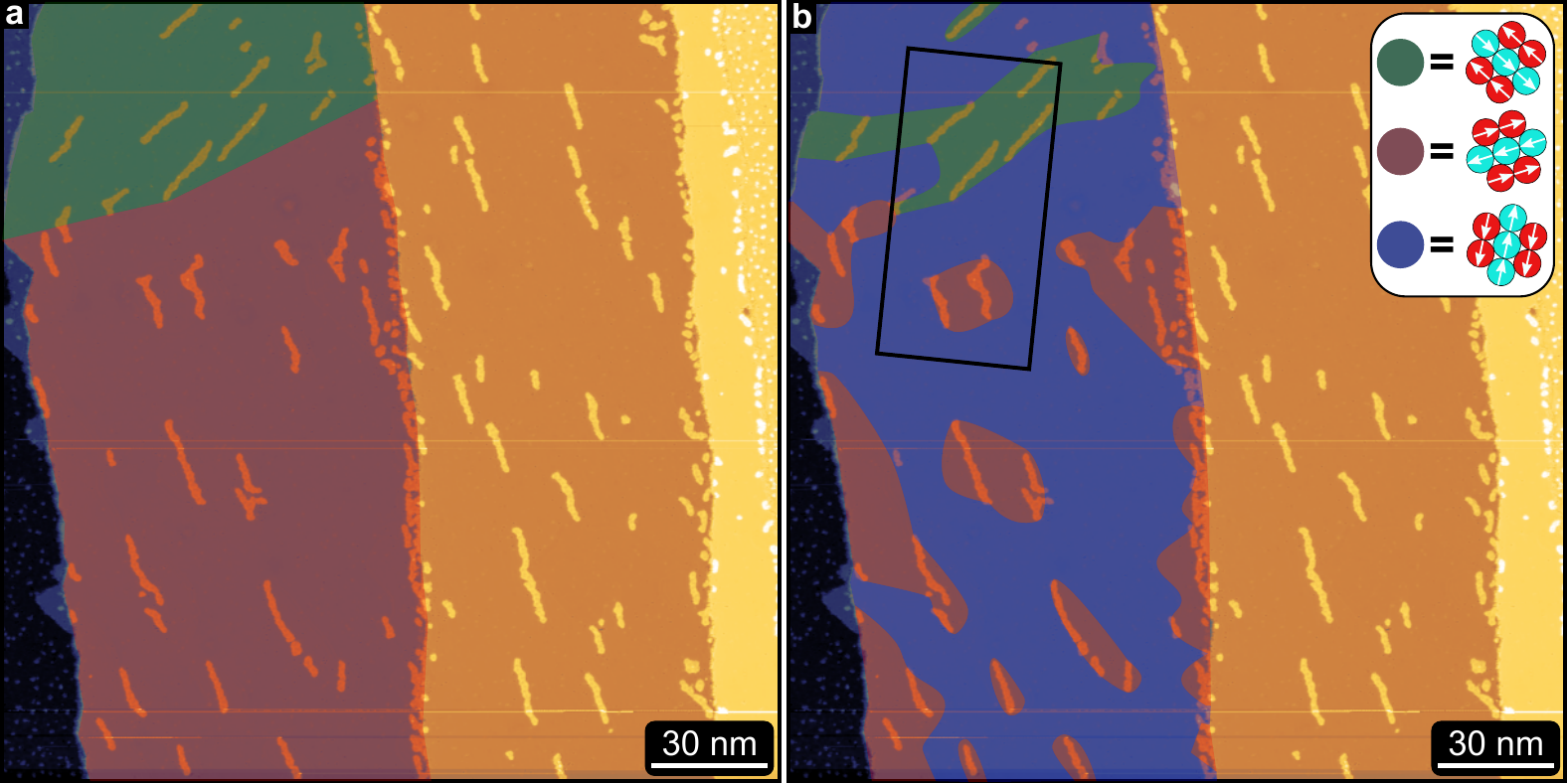}
    \caption{\textbf{| Co nanostructures pin magnetic domains. a,b,}~STM image of Co nanostructures deposited on Mn/Re(0001) at 80\,K. The magnetic state changed between growth and measurement. Red, green and purple tint indicate the orientational domains for the terrace on the left side of each scan frame present during growth and during imaging respectively. The black rectangle indicates the sample area displayed in Fig.\,\ref{fig4}c. $U=5$\,mV, $I=5$\,nA, $T=4.2$\,K. }
    \label{fig-Ex1}
\end{figure*}

\end{document}